\documentclass[twocolumn,showpacs,aps,prl,superscriptaddress]{revtex4}

\usepackage{graphicx}
\usepackage{dcolumn}
\usepackage{amsmath}
\usepackage{epsfig}

\input babarsym
\newcommand{\BABARPubYear}    {07}
\newcommand{\BABARPubNumber}  {069}

\newcommand{\SLACPubNumber} {13066}

\def\figurebox#1#2#3{%
    \def\arg{#3}%
    \ifx\arg\empty
    {\hfill\vbox{\hsize#2\hrule\hbox to #2{\vrule\hfill\vbox to #1{\hsize#2\vfill}\vrule}\hrule}\hfill}%
    \else
    {\hfill\epsfbox{#3}\hfill}%
    \fi}

\begin{document}
\preprint{\babar-PUB-\BABARPubYear/\BABARPubNumber} 
\preprint{SLAC-PUB-\SLACPubNumber} 

\begin{flushleft}
\babar-PUB-\BABARPubYear/\BABARPubNumber\\
SLAC-PUB-\SLACPubNumber\\
\end{flushleft}

\title{
\large \bf
Searches for the decays $\Bz\to\ell^\pm \tau^\mp$ and $\Bp\to\ellp\nu$ ($\ell = e,\mu$) using
hadronic tag reconstruction
}

%
\author{B.~Aubert}
\author{M.~Bona}
\author{Y.~Karyotakis}
\author{J.~P.~Lees}
\author{V.~Poireau}
\author{X.~Prudent}
\author{V.~Tisserand}
\author{A.~Zghiche}
\affiliation{Laboratoire de Physique des Particules, IN2P3/CNRS et Universit\'e de Savoie, F-74941 Annecy-Le-Vieux, France }
\author{J.~Garra~Tico}
\author{E.~Grauges}
\affiliation{Universitat de Barcelona, Facultat de Fisica, Departament ECM, E-08028 Barcelona, Spain }
% ...
\collaboration{The \babar\ Collaboration}
\noaffiliation

\date{\today}

\begin{abstract}
We present searches for the leptonic decays $\Bp\to\ellp\nu$ and the
lepton flavor violating decays $\Bz\to\ell^\pm\tau^\mp$, where $\ell = e,
\mu$, with data collected by the \babar\ experiment at SLAC.   
This search demonstrates a novel technique in which we fully reconstruct the accompanying 
\Bb in \upsbb events, and look for a monoenergetic lepton from the signal \B decay.
The signal yield is extracted from a fit to the signal lepton 
candidate momentum distribution in the signal \B rest frame.
Using a data sample of approximately 378 million \BB pairs ($342\,\mbox{fb}^{-1}$), we find 
no evidence of signal in any of the decay modes. Branching fraction upper limits of
$\mathcal{B}(\Bp\to\ep\nu)<5.2\times 10^{-6}$, $\mathcal{B}(\Bp\to\mup\nu)<5.6\times 10^{-6}$, 
$\mathcal{B}(\Bz\to\ep\taum)<2.8\times 10^{-5}$ and $\mathcal{B}(\Bz\to\mup\taum)<2.2\times 10^{-5}$,
are obtained at 90\% confidence level. 
\end{abstract}

\maketitle

In this paper, we present searches for the decays $\Bp\to\ellp\nu$ and the lepton flavor 
violating decays $\Bz\to\ell^\pm\tau^\mp$, where $\ell = e,\mu$~\cite{ref:charge}, using 
a technique in which the accompanying \B in the event is exclusively reconstructed. This method 
has not previously been used for searches for these modes and, although statistically limited
with the present \babar\ data sample, shows promise for future studies at, for example, 
a high luminosity Super B factory~\cite{ref:superB_CDR}.
While the former decay modes are allowed in the Standard Model (SM) and the latter are not, both 
are potentially sensitive to New Physics (NP) effects, such as contributions by neutral 
and charged non-SM Higgs~\cite{ref:NPmodels, ref:seesaw}.  

Searches for rare \B decays with neutrinos in the final state are challenging due 
to the limited availability of kinematic constraints. However, purely 
leptonic \B decays involving an electron or a muon have a clear experimental signature
in the form of a high momentum lepton. Combined with clean theoretical predictions 
due to the lack of QCD contributions in the final state, such leptonic \B decays present 
an ideal place to test the SM against NP models. 

In the SM, $\Bp\to\ellp\nu$ decays proceed via an annihilation of
\bbar and \u quarks into a virtual $W^+$ boson.
In the SM the branching fraction for this type of decay is given by~\cite{ref:smbranch}:
\begin{equation}
        \mathcal{B}(B^+\rightarrow \ellp\nul) =
        \frac{G_F^2 m_B m_l^2}{8\pi}\left(1-\frac{m_l^2}{m_B^2}\right)
        f_B^2 |V_{ub}|^2 \tau_B
	\label{eq:SMBranch}
\end{equation}
where $G_F$ is the Fermi coupling constant, $m_l$ is the lepton mass
and $m_B$, $\tau_B$ and $f_B$ are the mass, lifetime and decay constant for the \B meson. 
$\Vub$ is the Cabibbo-Kobayashi-Maskawa matrix element which describes the transition from $b$ to $u$ quarks~\cite{ref:lattice}. 
Within the SM, a determination of any one of the leptonic branching fractions represents a determination
of the product $|V_{ub}|\cdot f_B$, which can be directly compared with determinations from lattice calculations~\cite{ref:lattice},
$B$-mixing and semileptonic decay measurements~\cite{ref:BBmixing, ref:semilept}. 
As seen in Eq.(\ref{eq:SMBranch}), the decay rates are proportional
to ${m_l}^2$, resulting in SM predictions for the $\mu$ and $\electron$ modes 
which are suppressed by factors on the order of $250$ and $10^7$, respectively, compared with the $\tau$ mode.  Taking
the branching fraction $\mathcal{B} (\Bp \to \taup \nu_{\tau})=(1.31\pm0.48)\times10^{-4}$ from the combination of recent $\babar$ 
and BELLE results~\cite{ref:taunuBabar, ref:taunuBelle} implies $\mathcal{B}_{\rm SM} (\Bp\to\mup\num) \sim 5.2\times10^{-7}$ 
and $\mathcal{B}_{\rm SM} (\Bp\to\ep\nue) \sim 1.2\times10^{-11}$. New Physics contributions to these 
processes can enhance or suppress the decay rates compared to the SM, and may either preserve or violate
the relative rates of the three leptonic modes depending on the particular NP model~\cite{ref:NPmodels, ref:NPmodels2}. 
Thus, the $\electron$ and $\mu$ modes become particularily interesting in light of recent evidence 
for the \Bp\to\taup\nut decay mode. 
Currently, the most stringent published limits on $\Bp\to\ellp\nu$ are from the BELLE collaboration
with ${\mathcal B}(\Bp\to\ep\nu)<9.8\times 10^{-7}$ and ${\mathcal B}(\Bp\to\mup\nu) < 1.7 \times 10^{-6}$~\cite{ref:belleblnu}.
Earlier studies by CLEO and \babar\ collaborations are also available~\cite{ref:enu,ref:munu}. 

The lepton-flavor-violating (LFV) leptonic B decays, such as \Bz\to\ellp\taum,
are forbidden in the SM in the absence of non-zero neutrino masses, but can occur via one-loop
diagrams if neutrino oscillations are included. The rates of such processes, however, would be
substantially below current or anticipated future experimental sensitivities. On the other hand, many models of 
physics beyond the SM, in particular supersymmetric seesaw models~\cite{ref:seesaw}, predict dramatically 
higher rates for these decays.  In the case of Higgs-mediated LFV processes, couplings to heavier leptons are favored, 
making \Bz\to\ellp\taum particularily interesting. In the general flavor-universal MSSM, 
the branching fractions allowed for $\Bz\to\ellp\taum$ are $\sim 2\times10^{-10}$ \cite{ref:seesaw}. 
Such decays could be within the reach of a Super B factory with a data sample of 10 to 50$\,{\rm ab^{-1}}$.
The current best experimental limits on the branching fractions for these two decays are
$\mathcal{B}(\Bz\to\ep\taum)<1.1\times 10^{-4}$ and $\mathcal{B}(\Bz\to\mup\taum)<3.8\times 10^{-5}$,
set by the CLEO collaboration with 10$\,{\rm fb^{-1}}$ of data \cite{ref:elltau}.

The searches described in this work are based on a data sample of approximately 378 million 
\BB pairs, corresponding to an integrated luminosity of $342\,\mbox{fb}^{-1}$ collected at the \FourS 
resonance by the \babar\ detector at the \pep2\ asymmetric \epem\ storage ring.
Reconstructing the accompanying \B meson in specific hadronic modes 
prior to the signal selection allows the missing momentum vector of the neutrino(s) to be fully 
determined. The resulting increase in the energy resolution and the ability to infer the signal 
\B meson rest frame provide the extra kinematic handles that permit signal events to be cleanly 
distinguished from the background. 
Previous \B factory searches for $\Bp\to\ellp\nu$ and \Bz\to\ellp\taum have used an inclusive 
method in which the accompanying $\B$ is not explicitly reconstructed. This results in a significantly higher 
efficiency, but also a substantially increased background compared with the exclusive reconstruction 
method presented here. With the current level of luminosity, the inclusive method provides more 
stringent limits. However, due to the very low background achievable with the exclusive method, 
the two methods have about the same sensitivity for a statistically significant observation.
The method described in this paper will be the preferred approach for the high-precision 
studies of leptonic \B decays.

Charged-particle tracking and $dE/dx$ measurements for particle identification are provided by a five-layer
double-sided silicon vertex tracker and a 40-layer drift chamber contained within the magnetic field of a
$1.5\,\mbox{T}$ superconducting solenoid. A ring-imaging Cherenkov detector provides efficient particle
identification. The energies of neutral particles are measured with an electromagnetic calorimeter (EMC) consisting 
of 6580 CsI(Tl) crystals arrayed in a cylindrical barrel and in a forward endcap. 
Muon identification is provided by resistive plate chambers (partially replaced by limited streamer tubes for
a subset of the data that is used in this analysis) interleaved with the passive material comprising the
solenoid magnetic flux return. Signal efficiencies and background rates are estimated using a Monte Carlo (MC) 
simulation of the \babar\ detector based on GEANT4 \cite{ref:geant}. The \babar\ detector is described in 
detail in Ref.~\cite{ref:babar}.

Reconstructed charged tracks are assigned a particle hypothesis based on information from detector subsystems. 
$K_s^0$ candidates are selected by combining oppositely charged $\pi$
candidates and requiring that the $\pip \pim$ invariant mass satisfies $0.47\gevcc< m_{\pip \pim} <0.52\gevcc$.
$\pi^0$ candidates are obtained from the combination of EMC clusters with no associated tracks, each with 
a \FourS center-of-mass (CM) rest frame energy greater than 20\mev, for which the $\gamma \gamma$ 
invariant mass satisfies $ 115\mevcc < m_{\gamma \gamma} < 150\mevcc$.

Over 96$\%$ of the time, the \FourS resonance decays into a pair of \B mesons~\cite{ref:PDG2006}. 
Since the CM energy is precisely known at PEP-II, exclusive reconstruction of one of the two \B mesons, 
which we denote $\B_{\rm tag}$, fully determines the momentum four-vector of the other \B meson in the event.
Charged and neutral \B meson candidates are reconstructed in hadronic final states of the form 
$\B \to D^{(*)} X_{\rm had}$.  The reconstruction procedure begins with a $D^{(*)0}$ or $D^{(*)\pm}$ seed, to which
charged and neutral pions and kaons (which form the $\X_{\rm had}$ system) are then added.   
The combination of the $D^{(*)}$ and $\X_{\rm had}$ with the lowest value of $ \Delta E = |E_B - E_{\rm beam}|$ that satisfies the
condition $\Delta E < 0.2 \gev$ is chosen as the $\B_{\rm tag}$ candidate, where $E_B$ is the energy of the reconstructed 
\B meson and $E_{\rm beam}$ is the beam energy, both evaluated in the CM frame.  
We reconstruct \Dstarp in the \Dp\piz and \Dz\pip channels, and \Dstarz in the \Dz~\piz and \Dz$\gamma$ channels. 
The \Dp is reconstructed in the modes \Km\pip\pip, $K^{0}_{s}\pip$, $K^{0}_{s}\pip\piz$, 
\Km\pip\pip\piz and $K^{0}_{s}\pip\pip\pim$. For \Dz we consider the modes \Km\pip, 
\Km\pip\piz, \Km\pip\pip\pim and $K^{0}_{s}\pip\pim$.

Although multiple $D^{(*)} X_{\rm had}$ combinations may be present in a single event, this procedure permits, 
at most, a single $B_{\rm tag}$ candidate to be retained in any given event. For the $B_{\rm tag}$ candidate, 
we define the energy substituted mass, $m_{\rm ES} = \sqrt{E_{\rm beam}^2 - {\vec{p}_B}^{~2}}$, where $\vec{p}_B$ is 
the momentum of the $B_{\rm tag}$ candidate in the CM frame. $B_{\rm tag}$ candidates that are 
correctly reconstructed peak in $m_{\rm ES}$ near the nominal \B meson mass, while 
incorrectly reconstructed $\B_{\rm tag}$ candidates produce a combinatorial distribution. 
The signal events are required to lie within the range $5.270\gevcc<m_{\rm ES}<5.288\gevcc$. 
This reconstruction procedure results in a yield of approximately 2500 (2000) correctly reconstructed 
$\Bpm$ ($\Bz$) candidates per fb$^{-1}$ of data. 

Because the two \B mesons are produced with very little momentum in the CM frame, $\BB$ events typically produce
a more isotropic distribution of particles in the detector than non-resonant (``continuum'') backgrounds.  
Such backgrounds ($e^+e^- \to f \bar{f}$, where $f$ represents $u, d, s, c$ or any charged lepton) 
are suppressed by requiring $R_2<0.5$, where $R_2$ is the ratio of the second to the zeroth Fox-Wolfram moment \cite{ref:wolfram}
computed using all charged and neutral particles in the event. Further suppression is achieved by requiring 
$|\cos \theta_T|<0.90$, where $\theta_T$ is the angle between two thrust axes in the CM frame, the first computed 
using the particles from the $\B_{\rm tag}$, and the second using all other particles in the event. 

All particles that are not used in the $\B_{\rm tag}$ reconstruction are considered candidates
to be included in the reconstruction of the signal B meson. Since the CM energy is precisely known, reconstruction 
of the $B_{\rm tag}$ fully determines the $\B_{\rm signal}$ 4-vector. This permits the 2-body kinematics of the signal
decays to be exploited. In particular, these decays are expected to contain an electron or a muon with a
momentum $p^*$, in the $\B_{\rm signal}$ rest frame, of about $2.64\gevc$ ($2.34$\gevc) 
for the $\Bp\to\ellp\nu$ ($\Bz\to\ellp\taum$) channels, very close to the kinematic endpoint for \B decays.
 
Signal candidate events are initially selected by requiring the highest momentum track in the event 
(excluding tracks from the $\B_{\rm tag}$ reconstruction) to have a momentum of $1.7\gevc < p^* < 3.0\gevc$ 
and to satisfy particle identification (PID) criteria for either an electron or a muon. In events 
with a charged $\B_{\rm tag}$, the charge of the track is required to be opposite that of the $\B_{\rm tag}$, 
while for a neutral $\B_{\rm tag}$ the high-$p^*$ lepton is permitted to have either positive or negative charge. 

Once the $\B_{\rm tag}$ and the signal lepton candidate are identified, $\Bp \to \ellp \nu$ events should ideally 
have no other particles in the detector, while $\Bz \to \ellp \taum $ events should additionally contain only the $\taum$
decay daughters. For the latter, the \taum rest frame is calculated from the observed signal lepton, 
assuming the nominal energy and momentum of the $\taum$ for a 2-body \Bz decay. The six $\tau$ decay modes 
considered are listed in Table~\ref{tab:taudecays}. The second highest momentum track in the event (again, excluding 
$\B_{\rm tag}$ reconstruction) is assumed to be a $\tau$ daughter, and is required to have a charge opposite 
to the primary signal lepton. If this track satisfies electron or muon PID, the event is considered to be a 
leptonic $\tau$ decay. Otherwise, the track is assumed to be a pion and the quantity $\Delta E_{\tau}$ is calculated 
for the hadronic decay modes listed in Table~\ref{tab:taudecays}. $\Delta E_{\tau} = \sum E_{\pim,\piz} + p_{\nu} - m_\tau$, 
where $m_{\tau}=1.777\gevcc$, the sum is over the $\tau$ daughter candidates, the momentum of the neutrino is 
$p_{\nu}=|\sum\vec{p}_{\pim,\piz}|$, and all quantities are measured in the $\taum$ rest frame. We assign the decay 
mode for which $|\Delta E_{\tau}|$ is smallest, 
requiring additional conditions for the decay modes that proceed through the intermediate resonances 
$\rho^{-}\to\pim\piz$, ${a_1}^{-}\to\pim\piz\piz$ and ${a_1}^{-}\to\pim\pim\pip$. We calculate the quantity 
$\cos \theta_{\tau-\rho} = (2E_{\tau}E_{\rho}-m_{\tau}^2-m_{\rho}^2) / (2|\vec{p_{\tau}}||\vec{p_{\rho}}|)$,
where ($E_{\tau},\vec{p}_{\tau}$) and ($E_{\rho},\vec{p}_{\rho}$) are the four-momenta in the 
$\B_{\rm signal}$ frame, and $m_{\tau}$ and $m_{\rho}$ are the masses of the $\tau$ and $\rho$. 
For a correctly reconstructed $\rho$, this quantity peaks near unity. If the candidate does 
not satisfy $\cos \theta_{\tau-\rho}>0.70$ the mode with the next smallest $|\Delta E_{\tau}|$ (if one is present) 
is selected instead. Analogous quantities are calculated for the \taum\to\pim\piz\piz\nut and \taum\to\pim\pim\pip\nut modes, 
but with an $a_1^\pm$ instead of a $\rho^\pm$. The requirements of $\cos \theta_{\tau-a_1}>0.45$ 
and $\cos \theta_{\tau-a_1}>0.35$ are used for the two cases, respectively. There are no 
additional requirements on the $\rho$ or $a_{1}$.

\begin{table}[htbp]
\caption{The $\tau$ decays considered are listed with their branching fractions, in percent~\cite{ref:PDG2006}.}
\begin{center}
\begin{tabular}{c|c} \hline 
$\tau$ decay mode & Branching Fraction  \\ \hline
$e^{-}\nueb\nut$     & 17.84$\pm$0.05 \\ 
$\mu^{-}\numb\nut$   & 17.36$\pm$0.05 \\
\pim\nut              & 10.90$\pm$0.07  \\
\pim\piz\nut          & 25.50$\pm$0.10  \\
\pim\piz\piz\nut      & 9.25$\pm$0.12   \\
\pim\pim\pip\nut      & 9.33$\pm$0.08   \\ \hline 
\end{tabular}
\end{center}
\label{tab:taudecays}
\end{table}

Additional background, for both $\Bp\to\ellp\nu$ and $\Bz\to\ellp\taum$ decays,  
can arise from a variety of sources, including beam backgrounds, 
unassociated hadronic shower fragments, reconstruction artifacts, bremsstrahlung and 
photon conversions. We demand that events have no more than two extra charged tracks and 
six extra neutral clusters, allowing the presence of low energy particles not necessarily 
associated with the decay of the \FourS. Requirements on the missing 
momentum and extra energy in the event are utilized to ensure that such particles are 
unimportant for the analysis. Since many of the following requirements are optimized for 
each signal mode individually, we quote the approximate values.

The extra momentum in the event is represented by 
$\Delta P_{\rm miss} = | \vec{p}_{\rm miss} + \sum \vec{p}_{\ell,\pi} |$,
where $p_{\ell,\pi}$ are the momenta of the lepton or pion candidate(s) assumed to be
recoiling against the neutrinos. The missing momentum is calculated according to 
$\vec{p}_{\rm miss} = \vec{p}_{\FourS} - \vec{p}_{\B_{\rm tag}} - \vec{p}_{\rm all}$, 
where $\vec{p}_{\rm all}$ is the momentum of all tracks and clusters left after the 
$\B_{\rm tag}$ reconstruction. $\Delta P_{\rm miss}$ is calculated in the rest frame 
of the parent of the neutrino(s), so that the missing momentum balances the sum of 
other signal particles' momenta. The signal events are selected by requiring 
$\Delta P_{\rm miss}$ to be less than $0.5\gevc$.

For $\Bp\to\ellp\nu$ modes we also consider the direction of the missing
momentum $\cos {\theta_{p}}_{\rm miss} = {p_z}_{\rm miss} / p_{\rm miss}$,
where the subscript $z$ indicates the component of the momentum in the direction
parallel to the beam pipe, as measured in the \FourS CM frame. The requirement 
$-0.76<\cos \theta_{p_{\rm miss}}<0.92$ is determined by the geometry of the detector; events 
where $p_{\rm miss}$ points outside of the detector acceptance in the forward or backward direction 
are excluded. 

\begin{figure}[htbp]
\begin{center}
\includegraphics[height=3.0cm]{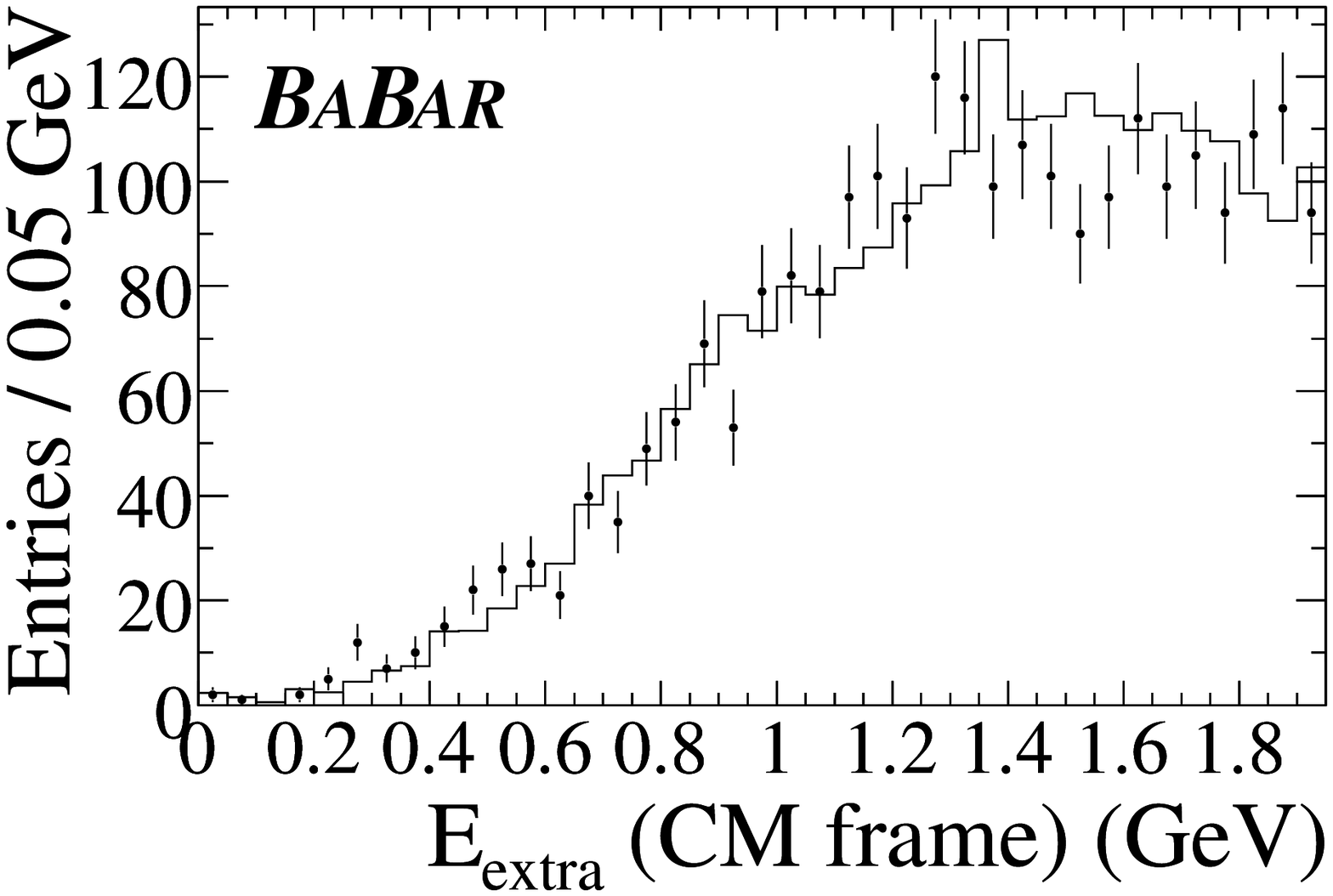}
\includegraphics[height=3.0cm]{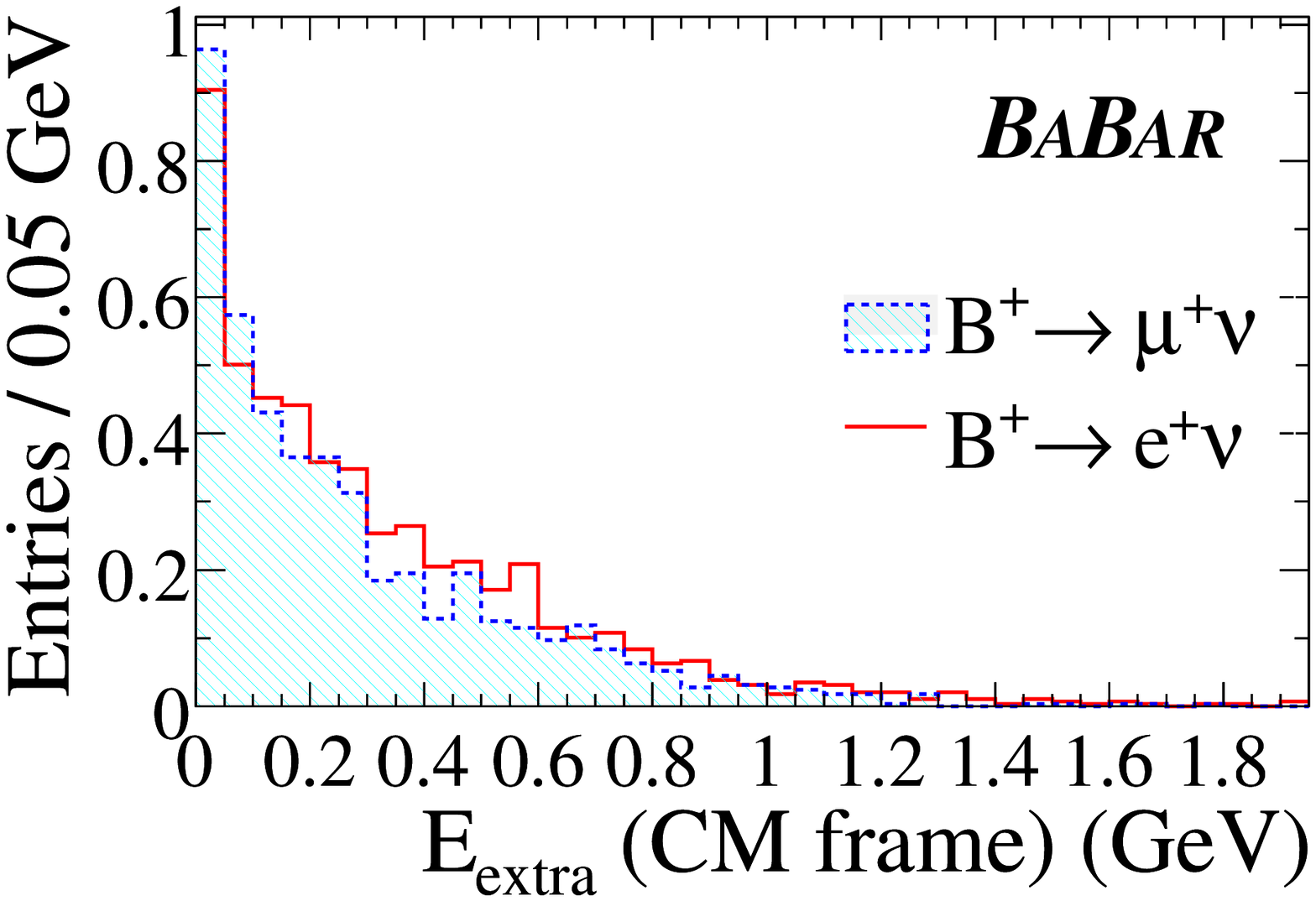}
\includegraphics[height=3.0cm]{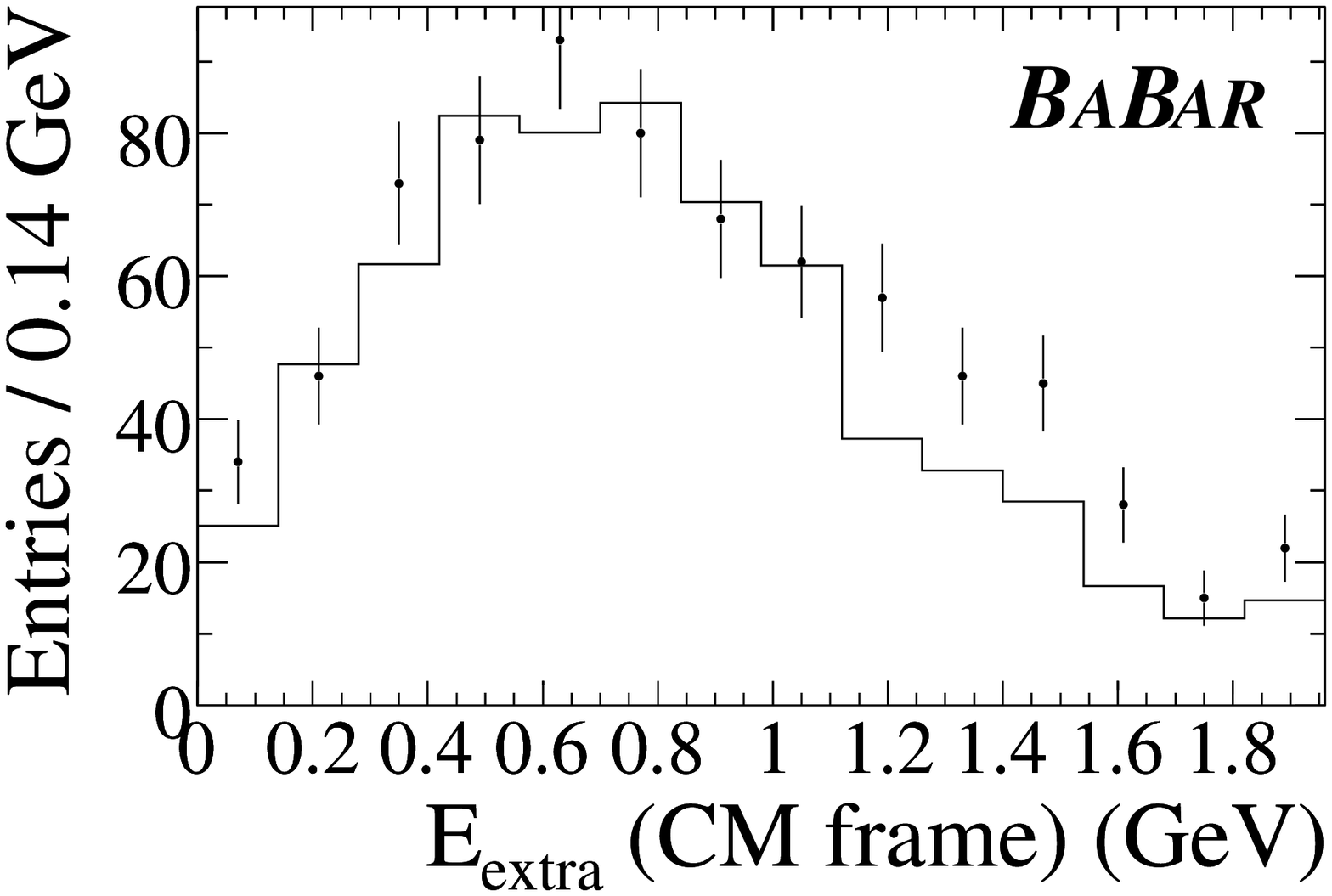}
\includegraphics[height=3.0cm]{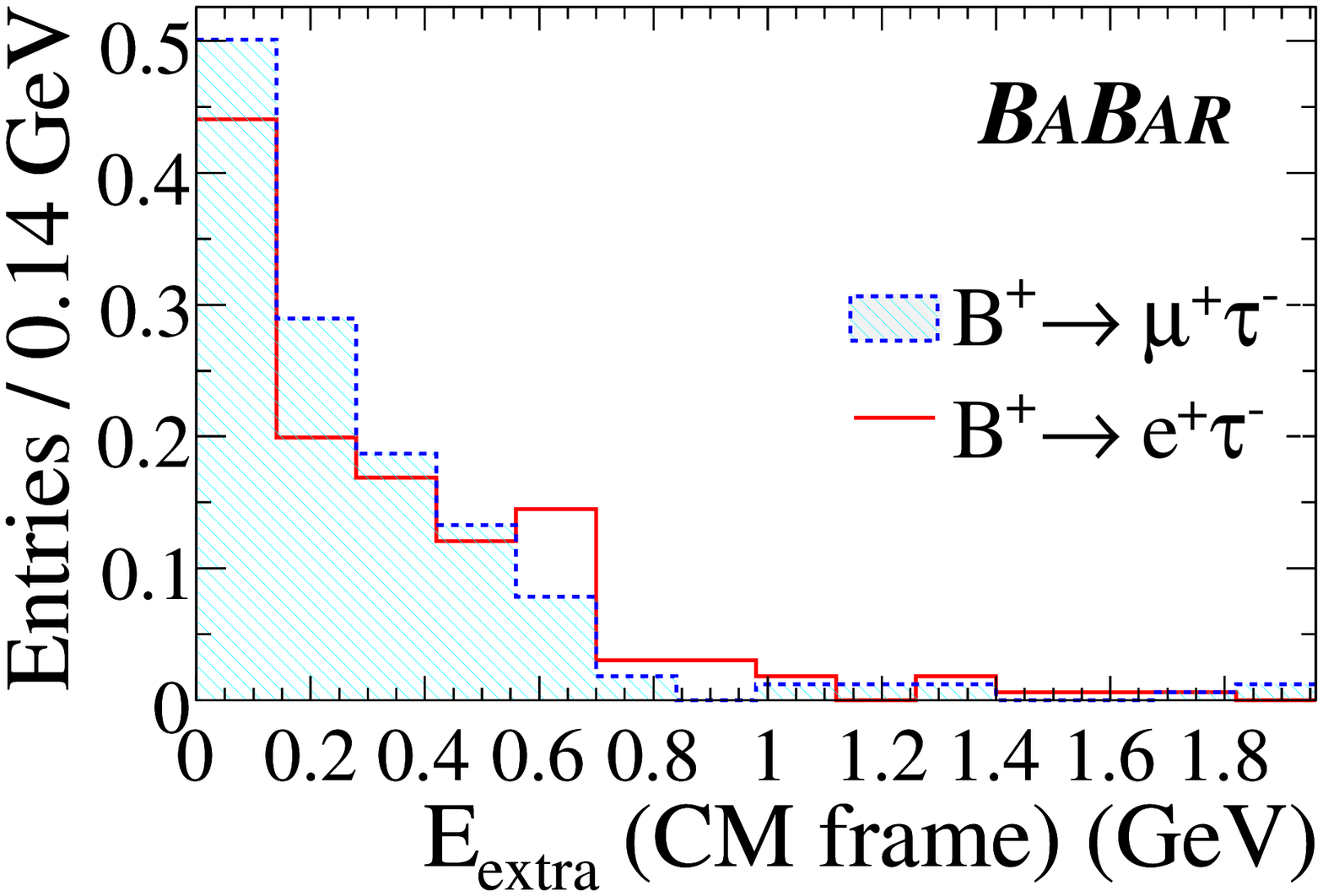}
\caption{$E_{\rm extra}$ distributions for the background simulation and data (left) and the signal (right) after
most of the selection criteria have been applied. The upper plots are for $\Bp\to\ellp\nu$ modes and the lower
plots are for \Bz\to\ellp\taum modes. The background distributions show electron and muon modes together, 
as they are nearly identical. The background is almost completely dominated by \BB events. 
The signal modes are shown with a branching fraction of $10^{-5}$.} 
\label{fig:Eextra}
\end{center}
\end{figure}

The quantity $E_{\rm extra} = \sum E_{\rm track} + \sum E_{\rm cluster} - E_{\ellp} - \sum E_{\ellm,\pim,\piz}$
describes the amount of energy recorded by the detector that is not accounted for by the high 
momentum lepton and $\taum$ daughters (in the case of \Bz\to\ellp\taum). The clusters and tracks associated with the 
reconstruction of $\B_{\rm tag}$ are excluded from the sums, and only clusters with energy 
more than 50\mev in the CM frame are considered. We require $E_{\rm extra}$ to be less than $1.0\gev$ in the 
CM frame. The signal and background distributions for $E_{\rm extra}$ are shown in Fig. \ref{fig:Eextra}.

\begin{figure}[htbp]
\begin{center}
\includegraphics[height=3cm]{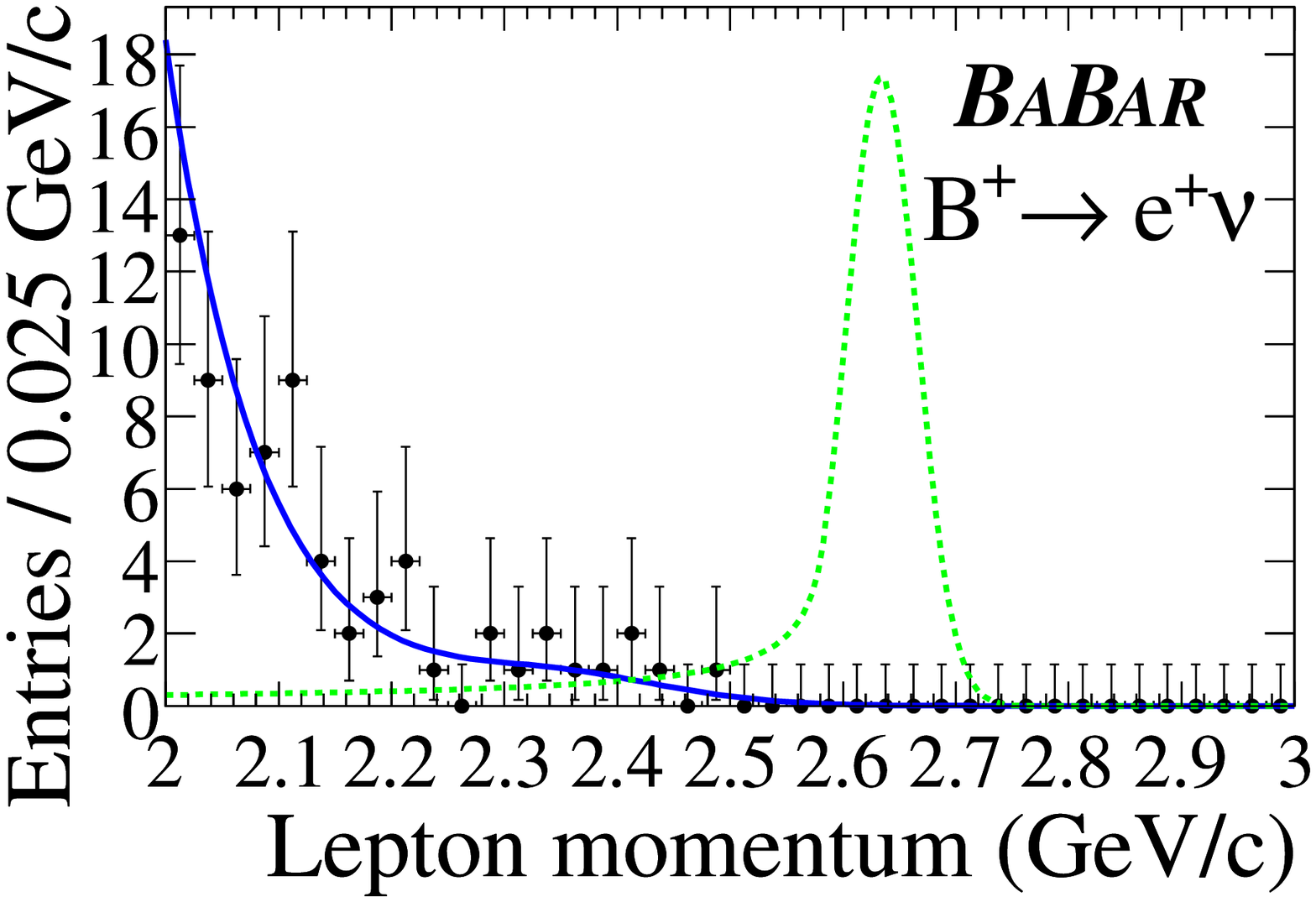}       
\includegraphics[height=3cm]{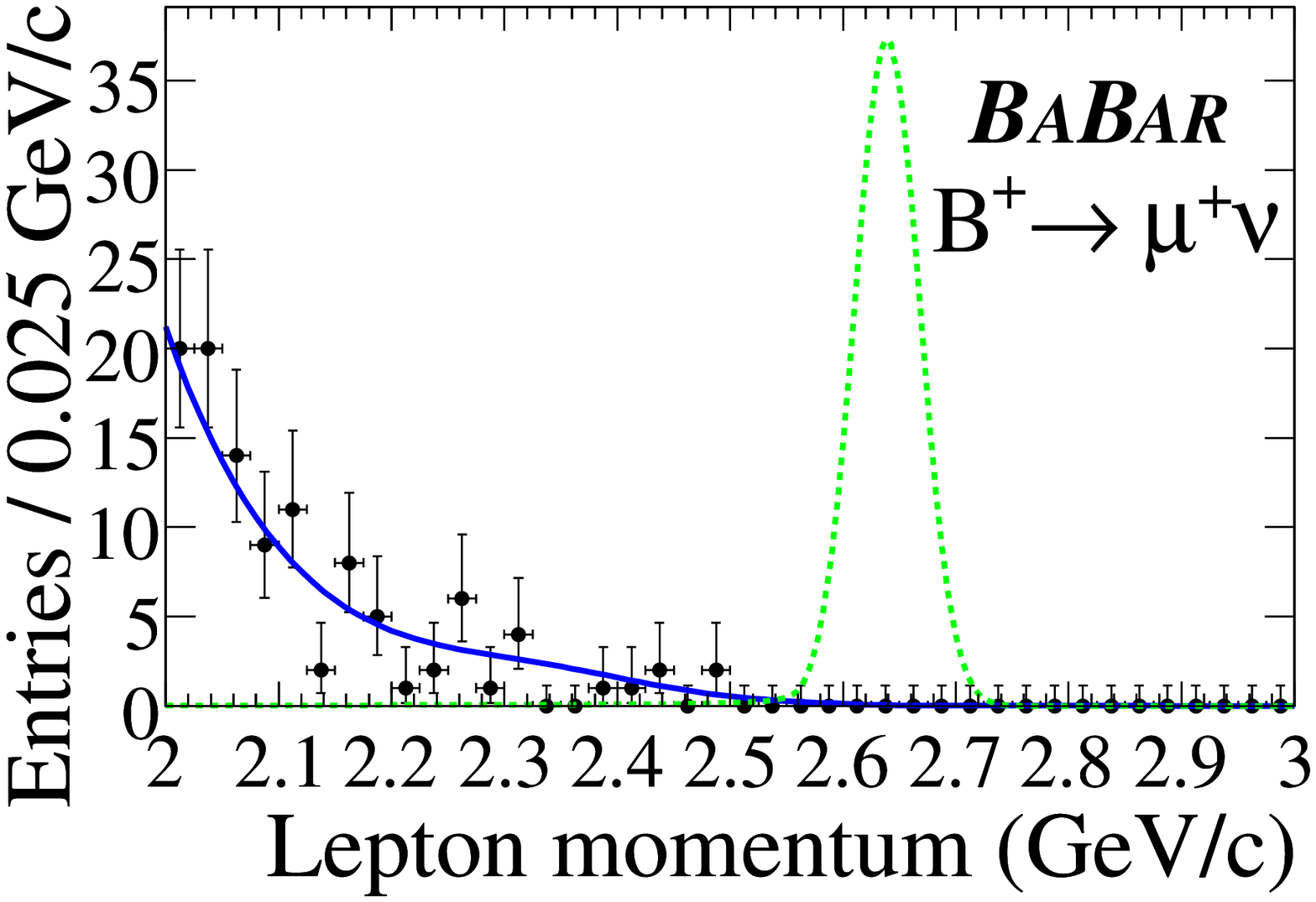}      
\includegraphics[height=3cm]{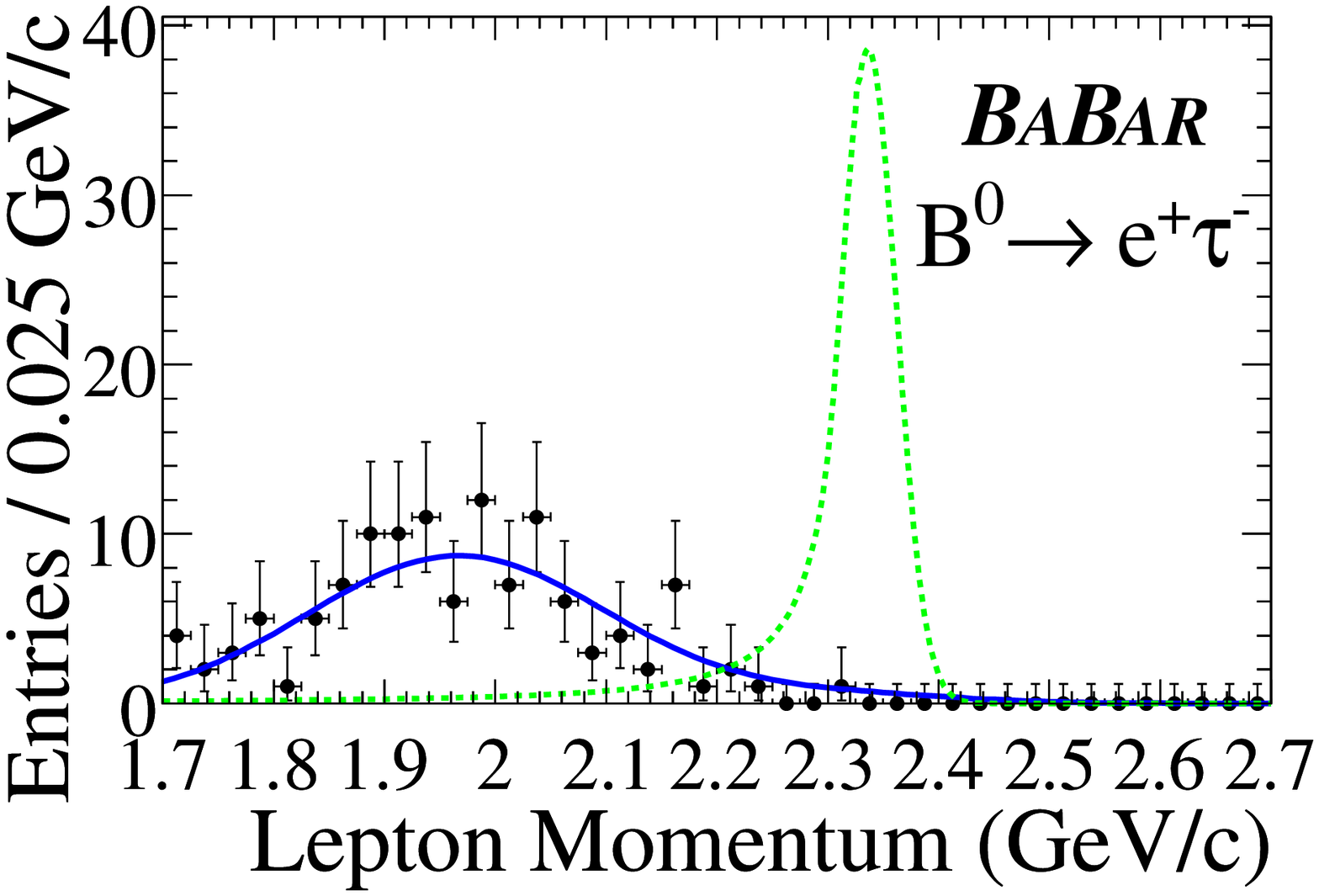}       
\includegraphics[height=3cm]{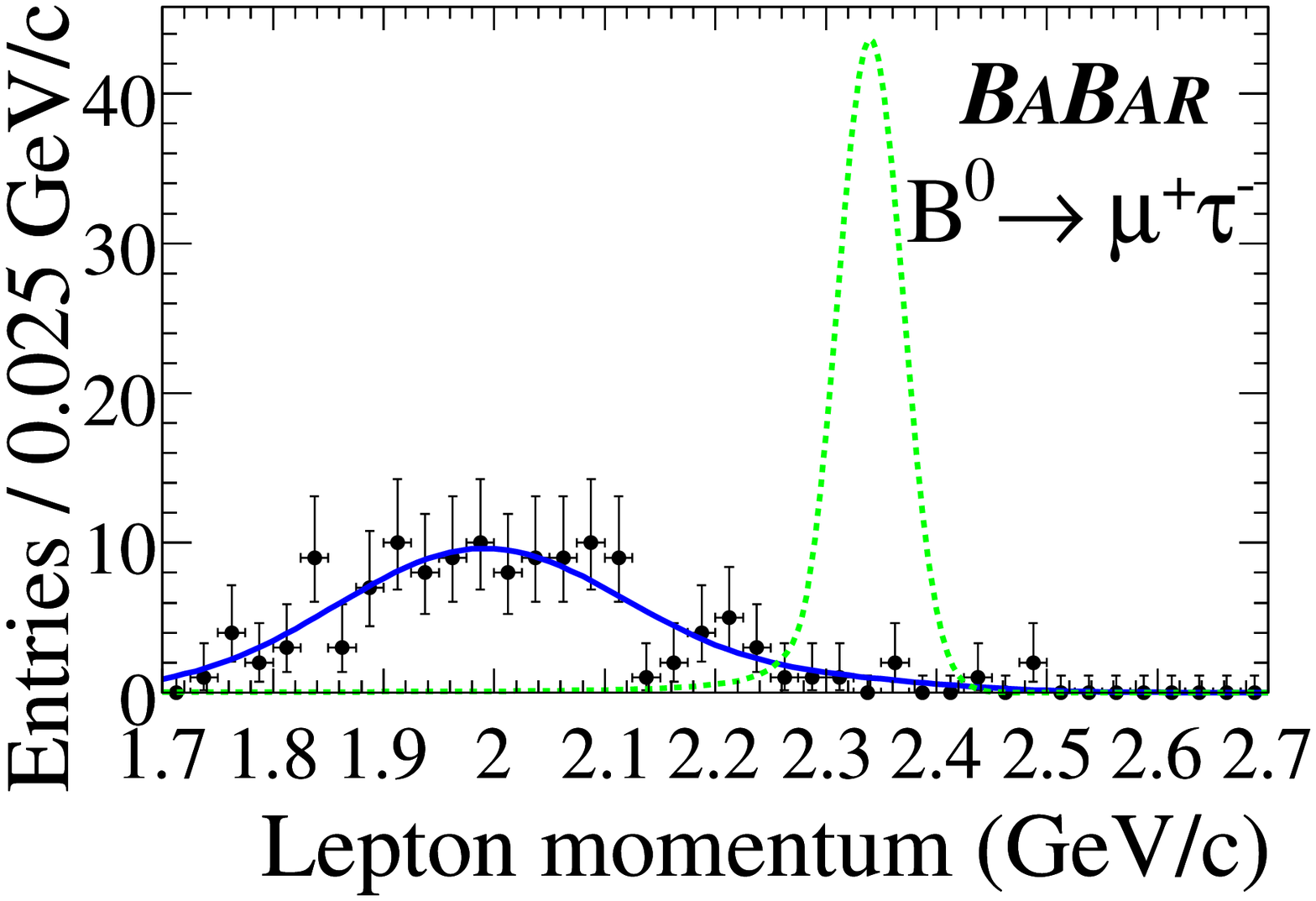}      
\caption{The unbinned maximum likelihood fits on the lepton momentum. The dashed line,
representing the signal PDF with an arbitrary scaling, indicates where the signal is expected.} 
\label{fig:fits}
\end{center}
\end{figure}

The signal yields are extracted from unbinned maximum likelihood fits to the signal 
lepton momentum distributions, as measured in the $\B_{\rm signal}$ frame. 
The signal and background MC distributions are fitted by phenomenological probability density functions (PDF). 
The signal distributions are modeled with Crystal Ball functions~\cite{ref:crystalball} to account 
for the energy loss due to unreconstructed bremsstrahlung photons. The $\Bp\to\ellp\nu$ background is modeled 
with an exponential decay and a Gaussian distribution, while the \Bz\to\ellp\taum background is 
modeled with a double Gaussian distribution. The PDF parameters are determined from simulated events. 
The fit is performed using the following likelihood function: 
\begin{equation}\label{eq:LHfunct}
\mathcal{L}(n_s, n_b)=\frac{e^{-(n_s+n_b)}}{N!}\prod_{i=1}^N[n_s f_s(i)+n_b f_b(i)],
\end{equation}
where $N$ is the total number of events in the fit region, $f_s(i)$ and $f_b(i)$ are the PDFs 
for the signal and background, and $n_b$ and $n_s$ are the number of background and signal events. 
All parameters of the signal and background PDFs remain fixed, while $n_s$ and $n_b$ are 
allowed to float. The fits are restricted to the ranges in $p^*$ shown in Fig.\ref{fig:fits}. 

The 90\% confidence level (C.L.) upper limit on the branching fraction $\mathcal{B}$ is determined by solving 
for $\mathcal{B}^{90\%}$ in 
$0.90 = \int_{0}^{\mathcal{B}^{90\%}} \mathcal{L}(\mathcal{B}) d\mathcal{B} / \int_{0}^{\infty} \mathcal{L}(\mathcal{B}) d\mathcal{B}$
for events lying in the signal regions of $2.40\gevc<p^{*}<2.75\gevc$ for $\Bp\to\ellp\nu$ and $2.20\gevc<p^*<2.42\gevc$ 
for $\Bz\to\ellp\taum$. $\mathcal{B}$ is related to the signal yield $n^*_s$ through a substitution 
$n^*_s=\epsilon_{\rm tot}\times 2\times N_{\rm \BB}\times \mathcal{B}$, where $\epsilon_{\rm tot}$ is the total signal 
selection efficiency and $N_{\rm \BB}$ is the number of \BpBm or \BzBzb pairs in the data sample. 
The signal selection efficiencies, expected number of background events and fit results are given 
in Table~\ref{tab:results}. The number of signal events given by the fits is consistent with zero for all decay modes. 
The uncertainties in Table~\ref{tab:results} are statistical except for those shown for $\mathcal{B}$ which are the 
statistical and systematic uncertainties added in quadrature.

\begin{table}[htbp]
\caption{Signal selection efficiency $\epsilon_{\rm tot}$ determined from MC, the fitted numbers of signal 
and background events in the signal regions $n^*_s$ and $n^*_b$, and the branching fractions $\mathcal{B}$. The uncertainties
for $\mathcal{B}$ include statistical and systematic terms. The uncertainties for the other quantities are 
statistical only.}
\begin{center}
\begin{tabular}{c|c c c c} \hline
& \multicolumn{4}{c}{Signal Mode} \\
			     	& $\ep\nu$ & $\mup\nu$ & $\ep\taum$ & $\mup\taum$  \\ \hline
$\epsilon_{\rm tot} \times10^5$ & $135\pm4$ & $120\pm4$  & $32\pm2$  & $27\pm2$    \\ 
$n^*_b$ MC                   & $2.66\pm0.13$  & $5.74\pm0.25$   & $8.69\pm0.27$   & $12.14\pm0.45$     \\
$n^*_b$                      & $2.67\pm0.19$  & $5.67\pm0.34$   & $9.35\pm0.35$   & $13.03\pm0.31$     \\
$n^*_s$                      & $-0.07\pm0.03$ & $-0.11\pm0.05$   & $0.02\pm0.01$   & $0.01\pm0.01$      \\ 
$\mathcal{B}\times10^{-6}$ & $-0.1^{+2.6}_{-1.7}$ & $-0.2^{+2.7}_{-1.8}$ & $0^{+15}_{-10}$ & $0^{+11}_{-7}$  \\ \hline
$\mathcal{B}^{90\%~C.L.}$    & $5.2\times10^{-6}$ & $5.6\times10^{-6}$ & $2.8\times10^{-5}$ & $2.2\times10^{-5}$  \\ \hline
\end{tabular}
\end{center}
\label{tab:results}
\end{table}

The systematic uncertainties arising from the fitting procedure are studied by repeating the 
procedure on additional simulated samples, generated according to the PDFs, with varying number of signal events. 
Systematic effects are studied by repeating the procedure with PDF parameters varied by their uncertainties. 
For the case of zero signal events, we find negligible effects on the branching fraction values, and
take the standard deviation of $n_s$ and $n_b$ from their expected values in the fits as systematic uncertainties. 
We find the fits to be well behaved and having no significant sources of bias, introducing
no additional uncertainties. Total uncertainties associated with the fitting procedure are listed 
in Table III for each decay mode. 

The discrepancies between simulation and data are treated as detailed in the following paragraphs.
The number of correctly reconstructed $\B_{\rm tag}$ events in the $m_{\rm ES}$ signal region 
is compared between simulation and data. The $m_{\rm ES}$ 
distributions for simulation and data are fitted with a combination of ARGUS~\cite{ref:argus} 
and Crystal Ball functions, allowing the number of $m_{\rm ES}$ peaking events to be estimated 
by integrating the peaking component between 5.270\gevcc and 5.288\gevcc. We find the simulation 
to underestimate the number of events with a good $\B_{\rm tag}$ and scale the signal selection efficiency 
by a factor of $1.11\pm0.06$ ($1.05\pm0.06$) for events with a neutral (charged) $B_{\rm tag}$.

In addition, the PID efficiencies in simulation are corrected for the 2-5\% lower efficiencies in data. 
We assign associated uncertainties of about $2\%$ for high momentum particles (signal lepton), and 
about $5\%$ for tau daughters.
The misidentification rate of leptons and pions is found to be negligible in the simulated samples, 
after all selection criteria are applied. An uncertainty in the tracking algorithm introduces 
an additional 0.8$\%$ systematic uncertainty for each charged track present in any given signal mode 
(e.g. $1.6\%$ for \Bz\to\ellp\taum, \taum\to\pim\nut). The uncertainties for \Bz\to\ellp\taum modes 
are calculated as weighted averages of all $\taum$ decay modes. 

Table \ref{tab:errors} lists the sources and the magnitudes of the uncertainties with their 
effect on $\mathcal{B}$. The uncertainties are incorporated into the final results by varying the 
branching fraction assumption by its uncertainty when integrating $\mathcal{L}$ for 
the 90\%~C.L. upper limit.

\begin{table}[htbp]
\caption{The sources and magnitudes of systematic uncertainties, in percent. 
}
\begin{center}
\begin{tabular}{c|c c c c } \hline
& \multicolumn{4}{c}{Signal Mode} \\
Uncertainty source    & \ep\taum     & \mup\taum     & $\ep\nu$    & $\mup\nu$    \\ \hline
Signal Fit            & 5.6          & 10.6          & 4.3         & 8.2             \\
Background Fit        & 3.9          & 3.1           & 5.1         & 7.8            \\  \hline
$B_{\rm tag}$ efficiency & 6.4       & 6.4           & 5.8         & 5.8             \\
PID efficiency        & 5.3          & 5.8           & 1.0         & 2.0              \\
MC Statistics         & 8.6          & 7.4           & 3.0         & 2.8             \\
Tracking efficiency   & 1.7          & 1.7           & 0.8         & 0.8             \\ \hline
$N_{\BB}$             & 1.1          & 1.1           & 1.1         & 1.1             \\ \hline 
\end{tabular}
\end{center}
\label{tab:errors}
\end{table}

We have presented searches for the rare leptonic decays $\Bp\to\ellp\nu$ and $\Bz\to\ell^\pm\tau^\mp$, 
where $\ell = e, \mu$, using a novel hadronic tag reconstruction technique. 
We find no evidence of signal in any of the decay modes in a data sample of approximately 378 million \BB pairs 
($342\,\mbox{fb}^{-1}$), and set the branching fraction upper limits at
$\mathcal{B}(\Bp\to\ep\nu)<5.2\times 10^{-6}$, $\mathcal{B}(\Bp\to\mup\nu)<5.6\times 10^{-6}$,
$\mathcal{B}(\Bz\to\ep\taum)<2.8\times 10^{-5}$ and $\mathcal{B}(\Bz\to\mup\taum)<2.2\times 10^{-5}$,
at 90\% confidence level. While these upper limits on $\mathcal{B}(\Bp\to\ep\nu)$ and $\mathcal{B}(\Bp\to\mup\nu)$
complement the more stringent limits available from inclusive studies~\cite{ref:belleblnu, ref:munu}, 
the $\Bz\to\ep\taum$ and $\Bz\to\mup\taum$ results are the most stringent published upper limits available.

We are grateful for the 
extraordinary contributions of our \pep2\ colleagues in
achieving the excellent luminosity and machine conditions
that have made this work possible.
The success of this project also relies critically on the 
expertise and dedication of the computing organizations that 
support \babar.
The collaborating institutions wish to thank 
SLAC for its support and the kind hospitality extended to them. 
This work is supported by the
US Department of Energy
and National Science Foundation, the
Natural Sciences and Engineering Research Council (Canada),
the Commissariat \`a l'Energie Atomique and
Institut National de Physique Nucl\'eaire et de Physique des Particules
(France), the
Bundesministerium f\"ur Bildung und Forschung and
Deutsche Forschungsgemeinschaft
(Germany), the
Istituto Nazionale di Fisica Nucleare (Italy),
the Foundation for Fundamental Research on Matter (The Netherlands),
the Research Council of Norway, the
Ministry of Science and Technology of the Russian Federation, 
Ministerio de Educaci\'on y Ciencia (Spain), and the
Science and Technology Facilities Council (United Kingdom).
Individuals have received support from 
the Marie-Curie IEF program (European Union) and
the A. P. Sloan Foundation.

\end{document}